\documentclass[a4paper]{jpconf}

\newcommand{\mq}{\mathbf{q}}
\newcommand{\mf}{\mathbf{f}}
\newcommand{\mpsi}{\boldsymbol{\psi}}

\newcommand{\mbpsi}{\bar{\boldsymbol{\psi}}}

\newcommand{\mD}{\mathbb{D}}
\newcommand{\mbD}{\overline{\mathbb{D}}}

\newcommand{\p}[1]{(\ref{#1})}

\newcommand{\cD}{{\cal D}}

\newcommand{\cE}{{\cal E}}

\newcommand{\bY}{{\overline Y}}
\newcommand{\bD}{{\overline D}}

\newcommand{\bQ}{{\overline Q}{}}
\newcommand{\bS}{{\overline S}{}}

\newcommand{\bpsi}{{\bar\psi}{}}

\newcommand{\bnabla}{{\overline \nabla}{}}

\newcommand{\be}{\begin{equation}}
\newcommand{\ee}{\end{equation}}
\newcommand{\bea}{\begin{eqnarray}}
\newcommand{\eea}{\end{eqnarray}}

\newcommand{\ba}{\begin{array}} \newcommand{\ea}{\end{array}}

\def\im{{\rm i}}

\newcommand{\nn}{\nonumber}

\usepackage{amscd,amsmath,amssymb}

\begin{document}
\title{$N=4$, $d=3$ Born-Infeld theory in component approach}
\author{N Kozyrev and S Krivonos}
\address{Bogoliubov  Laboratory for Theoretical Physics, JINR,
141980 Dubna, Russia}
\ead{nkozyrev,krivonos@theor.jinr.ru}

\begin{abstract}\noindent
Using the formalism of nonlinear realizations we construct the component on-shell action of the $N=4$, $d=3$ Born-Infeld theory, which is the action of $N=2$, $d=3$ vector supermultiplet, fixed by invariance with respect to the additional spontaneously broken $N=2$, $d=3$ supersymmetry. Our construction shows that dealing with the systems with partial breaking of supersymmetry with vector fields in the multiplet, it is preferrable to use their formulation in terms of fermionic superfields only.
\end{abstract}

\setcounter{page}{1}
\setcounter{equation}{0}

\section{Introduction}
The natural way to deal with the systems with spontaneous symmetry breaking is the formalism of nonlinear realizations as the appearing Goldstone fields can be identified with the parameters of coset space, which are associated with generators of broken symmetries \cite{phenoml1}, \cite{phenoml2}, \cite{phenoml3}. Within this approach, it is possible to systematically find the transformation laws with respect to all the symmetries in the game, the differential Cartan forms covariant with respect to these symmetries, and, sometimes, the invariant actions as integrals of products of forms.

This approach has been applied to systems with partial breaking of supersymmetry, with Goldstone fermionic superfields being associated with half of supercharges. Then it appears to be possible to find, in addition, the covariant relations between Goldstone superfields (inverse Higgs effect - \cite{invhiggs}), the irreducibility conditions on Goldstone multiplets and the covariant equations of motion by applying restrictions on some of the Cartan forms. While it is not possible to find the superfield action in this way (it is not an integral of Cartan form as the corresponding Lagrangian shifts under supersymmetry transformations), the construction of the component action is still possible. If one chooses the fermionic fields of the multiplet as the first components of the Goldstone fermionic superfields, they will be subjected to a simple transformation law with respect to the broken supersymmetry  (just the shifts), while other fields will remain invariant. It allows one to conclude that these fields enter into action only through covariantized measure and derivatives (as Volkov and Akulov found \cite{goldferm}), as well as the Wess-Zumino terms. The component action, therefore, can be found as a simple generalization of the bosonic one, covariantized with respect to the broken supersymmetry. A few remaining coefficients can then be fixed by the unbroken supersymmetry invariance.

This approach allowed the construction of a few PBGS systems in three and four dimensions, with chiral and hypermultiplets as Goldstone superfields \cite{D5membr}, \cite{D6brane}, \cite{D8brane}, with only scalar fields in the bosonic sector, as well as the $N=2$, $d=4$ Born-Infeld theory with the $N=1$, $d=4$ Goldstone vector multiplet \cite{N2d4BI}. All these constructions have their own difficulties. For theories with hypermultiplets, it is required to solve the highly nonlinear relations between the multiplet components. For the $N=2$, $d=4$ Born-Infeld theory it is necessary to construct the irreducibility conditions for vector supermultiplet that are both off-shell and covariant with respect to the broken supersymmetry. Therefore, it would be desirable to understand how to construct the theories that contain both scalar and vector fields in the multiplet. One of the simplest examples of such systems is the $N=4$, $d=3$ Born-Infeld theory, Lagrangian for which can be found by either dimensional reduction from $d=4$ or dualization of one scalar in the action of the membrane in $D=5$ \cite{D5membr}. It would be desirable, therefore, to construct it from first principles.

\section{The algebra, Cartan forms and derivatives}
The starting point of the construction of the action of the $N=4$, $d=3$ Born-Infeld theory is the algebra of its symmetries, which is just the $N=4$, $d=3$ Poincar\'e superalgebra. Its bosonic generators are Lorentz rotations $M_{ab} = - \big(M_{ab} \big)^\dagger $, translations $P_{ab} = \big(P_{ab} \big)^\dagger$ and single central charge $Z = Z^\dagger$. Their commutation relations are
\bea\label{N8d3Poincare1}
&&\big[ M_{ab}, M_{cd}   \big] =  \epsilon_{ac}M_{bd} + \epsilon_{ad} M_{bc}+ \epsilon_{bc}M_{ad} + \epsilon_{bd}M_{ac}, \nn \\
&&\big[ M_{ab}, P_{cd}   \big] =  \epsilon_{ac}P_{bd} + \epsilon_{ad} P_{bc}+ \epsilon_{bc}P_{ad} + \epsilon_{bd}P_{ac}. \eea
Optionally, an automorphism generator $K_{ab}$ could be added. With it, it would be just the bosonic algebra of the paper \cite{D4membr}.

This bosonic algebra is accompanied by the supercharges $Q_a$, $S_a$, $\bQ_a = \big( Q_a  \big)^\dagger$, $\bS_a = \big( S_a  \big)^\dagger$. (Anti)commutation relations involving them read
\bea\label{N8d3Poincare2}
&&\big\{  Q_a, \bQ_b  \big\} = \big\{  S_a, \bS_b  \big\} = 2 P_{ab}, \quad \big\{  Q_a, S_b  \big\} = 2 \epsilon_{ab}Z, \quad \big\{  \bQ_a, \bS_b  \big\} = 2 \epsilon_{ab}Z, \nn \\
&&\big[ M_{ab}, Q_c  \big] =  \epsilon_{ac} Q_b + \epsilon_{bc} Q_a, \quad \big[ M_{ab}, S_c  \big] = \epsilon_{ac} S_b + \epsilon_{bc} S_a.
\eea
The generator $K_{ab}$ mixes $Q_a$ with $\bS_a$ and $S_a$ with $\bQ_a$.

The coset element can be chosen as
\be\label{coset0}
g = e^{\im x^{ab}P_{ab}} e^{\theta^a Q_{a} + \bar\theta^a \bQ_a}e^{\im \mq Z} e^{\mpsi^a S_{a} + \mbpsi^a \bS_a},
\ee
with $M_{ab}$ and $K_{ab}$ generators in the stability subgroup. The transformations in this coset space are induced by left multiplication $g_0 g = g^\prime \, h$, where $h\sim e^{\gamma M+ \sigma K}$. Most important transformation laws of the coordinates and fields are those with respect to the unbroken and broken supersymmetries:
\bea\label{coset}
g_0 &=& e^{\epsilon^a Q_a + \bar\epsilon^a \bQ_a} \; \Rightarrow \; \delta_Q x^{ab} = \im \big( \epsilon^{(a}\bar\theta^{b)} +  \bar\epsilon^{(a}\theta^{b)} \big), \; \delta_Q \theta^a = \epsilon^a,  \\
g_0 &=&  e^{\varepsilon^a S_a + \bar\varepsilon^a \bS_a} \; \Rightarrow \; \delta_S x^{ab} = \im \big( \varepsilon^{(a}\mbpsi^{b)} +  \bar\varepsilon^{(a}\mpsi^{b)} \big), \; \delta_S \mpsi^a = \varepsilon^a, \; \delta_S \mq = \im \big(\varepsilon_a \theta^a + \bar\varepsilon_a \bar\theta^a  \big).\nn
\eea
They ensure that unbroken supersymmetry is realized in the superspace in standard way. Also the imply that $\mpsi_a$ is the Goldstone fermionic superfield, as expected.

The following differential forms are invariant with respect to the transformations in the coset:
\bea\label{Cartan1}
&&g^{-1}dg = \im \triangle x^{ab}P_{ab} + \im \triangle \mq Z + d\theta^a Q_a + d\bar\theta^a\bQ_a + d\mpsi^a S_a + d\mbpsi^a \bS_a,  \\
&&\triangle x^{ab} = dx^{ab} - \im \big( \theta^{(a}d\bar\theta^{b)} + \bar\theta^{(a}d\theta^{b)} + \mpsi^{(a} d\mbpsi^{b)} + \mbpsi^{(a} d\mpsi^{b)}\big), \;  \triangle \mq = d\mq -2\im \big( \mpsi_a d\theta^{a} + \mbpsi_a d\bar\theta^a   \big).\nn
\eea

$\triangle x^{ab}$, $d\theta^a$, $d\bar\theta^a$ can be used to define the useful covariant derivatives
\bea\label{covders}
&& dx^{ab}\,\partial_{ab} + d\theta^a \, \frac{\partial}{\partial \theta^a}+ d
\bar\theta^a \, \frac{\partial}{\partial \bar\theta^a} =\triangle x^{ab}\,\nabla_{ab} + d\theta^a \, \nabla_a + d
\bar\theta^a \, \bnabla_a \Rightarrow \nn \\
&&\nabla_{ab} = \Big( E^{-1} \Big)_{ab}{}^{cd} \partial_{cd}, \quad E _{ab}{}^{cd} = \delta_{(a}^c \delta_{b)}^d - \im \big( \mpsi^{(c}\partial_{ab}\mbpsi^{d)} +\mbpsi^{(c}\partial_{ab}\mpsi^{d)}   \big), \\
&&\nabla_a = D_a - \im \big( \mpsi^c \nabla_a \mbpsi^d + \mbpsi^c \nabla_a \mpsi^d   \big)\partial_{cd}, \quad \bnabla_a = \bD_a - \im \big( \mpsi^c \bnabla_a \mbpsi^d + \mbpsi^c \bnabla_a \mpsi^d   \big)\partial_{cd}, \nn
\eea
with
\be\label{covders2}
D_a = \frac{\partial}{\partial \theta^a} - \im \bar\theta^b \, \partial_{ab}, \; \bD_a = \frac{\partial}{\partial
\bar\theta^a} - \im \theta^b \, \partial_{ab}, \; \big\{ D_a, \bD_b  \big\} = -2\im \partial_{ab}.
\ee
These derivatives form an algebra
\bea\label{dercomm}
&&\big\{ \nabla_a, \nabla_b  \big\} = -2\im \big( \nabla_a \mpsi^c \, \nabla_b \mbpsi^{d} +   \nabla_b \mpsi^c \, \nabla_a \mbpsi^{d} \big)\nabla_{cd}, \nn \\
&&\big\{ \nabla_a, \bnabla_b  \big\} =-2\im \nabla_{ab} -2\im \big( \nabla_a \mpsi^c \, \bnabla_b \mbpsi^{d} +   \bnabla_b \mpsi^c \, \nabla_a \mbpsi^{d} \big)\nabla_{cd}, \nn \\
&&\big[ \nabla_{ab}, \nabla_c   \big] =-2\im \big( \nabla_{ab}\mpsi^m \, \nabla_c \mbpsi^{n} + \nabla_{ab}\mbpsi^m \, \nabla_c \mpsi^{n}    \big)\nabla_{mn}, \\
&&\big[ \nabla_{ab}, \nabla_{cd}  \big] = 2\im \big( \nabla_{ab}\mpsi^m \, \nabla_{cd}\mbpsi^n -  \nabla_{cd}\mpsi^m \, \nabla_{ab}\mbpsi^n  \big)\nabla_{mn}. \nn
\eea

Let us note that, as usual for such systems, one may define the covariant derivative $\cD_{ab}$, which acts on the components, as the $\theta \rightarrow 0$ limit of $\nabla_{ab}$:
\be\label{cE}
\cD_{ab} = \big( \cE^{-1} \big)_{ab}{}^{cd}\partial_{cd}, \;\;\; \cE_{(ab)}{}^{(cd)} = E_{ab}{}^{cd}|_{\theta\rightarrow 0} = \delta_{a}^{(c}\, \delta_{b}^{d)}- \im \big( \psi^{(c}\partial_{ab}\bpsi^{d)} +\bpsi^{(c}\partial_{ab}\psi^{d)}   \big),
\ee
where $\psi_a = \mpsi_a|_{\theta\rightarrow 0}$ is the first component of the fermionic Goldstone superfield.

\section{Irreducibility conditions}
As the Cartan forms \p{Cartan1} are invariant with respect to $P_{ab}$, $Z$, $Q_a$ and $S_a$ transformations, one may impose conditions, respected by all these symmetries, which express the $\mpsi_a$ and $\mbpsi_a$ in terms of the derivatives of the superfield $\mq$:
\be\label{omegaZ}
\triangle\mq |_{d\theta^a} =0 \; \Rightarrow \;\nabla_a \mq = -2\im \mpsi_a, \; \; \; \triangle\mq |_{d\bar\theta^a} =0 \; \Rightarrow \; \bnabla_a \mq = -2\im \mbpsi_a.
\ee
These constraints, however, do not imply any irreducibility conditions on the superfield $\mq$. These conditions should be postulated independently.

The ordinary $N=2$, $d=3$ vector multiplet can be described by the following irreducibility conditions (see e.g.\cite{Zup1}):
\be\label{3dvecirr1}
D^a D_a \mq =0,\;   \bD^a \bD_a \mq =0.
\ee
They leave in the superfield $\mq$ the following independent components: a real scalar field $q = \mq |_{\theta\rightarrow 0}$, two fermionic fields $\psi_a = \mpsi_a |_{\theta\rightarrow 0} $, $\bpsi_a = \mbpsi_a |_{\theta\rightarrow 0}$, as well as the auxiliary field $A= \big[D^{a},\bD_{a}\big]\mq|_{\theta\rightarrow 0}$ and a vector $V_{ab}=\im \big( \bD_b \mpsi_a + D_a \mbpsi_b \big) |_{\theta\rightarrow 0}$. The real part of $D_a \mbpsi_b$ reduces to a derivative of $\mq$:
\be\label{Dbpsireal}
D_a \mbpsi_b + \bD_b \mpsi_a  = \frac{\im}{2} \big\{ D_a, \bD_b  \big\}\mq  = \partial_{ab} \mq.
\ee
The vector $V_{ab}$ satisfies the Bianchi identity as a consequence of \p{3dvecirr1}, and can be, therefore, identified with the electromagnetic field strength. This can be most easily shown by acting by $D^2$ on the constraint $\bD^2\mq =0$ and commuting $D^2$ to the right:
\be\label{flatBI}
0 = D^2 \bD^2 \mq  \sim \partial_{ab}[D^a, \bD^b ]\mq\; \Rightarrow \; \partial_{ab}V^{ab} =0.
\ee

The most natural way to find the constraints covariant with respect to the broken supersymmetry is just to replace the $D_a$-derivatives by proper covariant ones
\be\label{3dvecirr2}
D^a D_a \mq =0 \; \rightarrow\; \nabla^a\nabla_a \mq =0, \;\;\; \bD^a \bD_a \mq =0 \;\rightarrow \; \bnabla^a \bnabla_a \mq =0.
\ee
Note that, as ordinary derivatives $D_a$ anticommute, the flat condition $D^a D_a \mq =0$ implies $D_a D_b \mq =0$. While the anticommutator of $\nabla_a$-derivatives is nontrivial, one may note that $U_{ab} = \nabla_{(a}\mpsi_{b)}$ satisfies the homogeneous equation as a result of the condition $\nabla^a\nabla_a \mq =0$:
\be\label{3dvecirr3}
-2\im U_{ab} = \frac{1}{2}\big\{ \nabla_a,\nabla_b \big\}\mq = -\im \big( U_{ac}\, \nabla_{b}\mbpsi_d + U_{bc}\, \nabla_a\mbpsi_d   \big)\nabla^{cd}\mq.
\ee
As the matrix of this equation is nondegenerate (it is just the unitary matrix in the lowest approximation, and $\nabla_{b}\mbpsi_d$ should not be singular in $\nabla_{ab}\mq$), one may conclude that $U_{ab} =0$ and present the constraints \p{3dvecirr2} as
\be\label{3dvecirr4}
\nabla_a \mpsi_b =0, \; \bnabla_{a}\mbpsi_b =0.
\ee
This is equivalent to putting to zero the $d\theta^a$, $d\bar\theta^a$ projections of the $d\mpsi_b$ $d\mbpsi_b$ Cartan forms \p{Cartan1}, respectively. On these constraints, obviously, $\big\{ \nabla_a, \nabla_b \big\}=0$.

Relation \p{Dbpsireal} has an analog in the nonlinear theory
\be\label{nablabpsireal}
\nabla_a \mbpsi_b + \bnabla_b\mpsi_a  = \frac{\im}{2}\big\{  \nabla_a, \bnabla_b  \big\}\mq.
\ee
Then one may substitute here the explicit form of the anticommutator, \p{dercomm} and the on-shell ansatz for $\nabla_a \mbpsi_b$,
\be\label{XBV}
\nabla_a \mbpsi_b  |_{\theta\rightarrow 0}= \frac{1}{2}\big( X_{(ab)} +\im V_{(ab)}\big),\;\; \bnabla_b \mpsi_a |_{\theta\rightarrow 0} = \frac{1}{2}\big( X_{(ab)} - \im V_{(ab)}\big).
\ee
Let us note that according to this ansatz $\nabla_a \mbpsi^a=0$, $\bnabla_a \mpsi^a=0$, as the auxiliary field should not appear. Equation \p{nablabpsireal} in the limit $\theta \rightarrow 0$ implies
\be\label{XBV4}
X_{ab} = \big( 1 - \frac{1}{8}V_{cd}V^{cd} - \frac{1}{8}X_{cd}X^{cd}  \big)\cD_{ab} q + \frac{1}{4}\big( X^{cd} \cD_{cd} q  \big) X_{ab} + \big( V^{cd} \cD_{cd} q  \big) V_{ab}.
\ee

Equation  \p{XBV4} is actually a source of the significant problem: while its solution can be found analytically, it appears to be too complicated to be practically useful. More difficulties come as one tries to calculate the Bianchi identity simplifying the expression $\nabla^2 \bnabla{}^2 \mq =0$. In the bosonic limit, the result is
\bea\label{BIid5}
&&M_{ab}{}^{cd} \partial_{cd}V^{ab} - \frac{1}{2} M_{ab}{}^{cd} \partial_{mn}q \big( \partial_{cd}V^{am}\, X^{bn} - \partial_{cd}X^{bn}\, V^{am}   \big) + \nn \\
&&+ \frac{1}{8}V_a^c\, X^{ad}\,\partial_{cd}X^{km}\, X^n_k \, \partial_{mn}q + \frac{1}{8} V_a^c X^{ad}\partial_{cd}V^{km}V_k^n \partial_{mn}q=0, \\
&&\mbox{where} \; M_{ab}{}^{cd} = \delta_{a}^{(c} \delta_{b}^{d)} + \frac{1}{4}X_a^{(c} X_b^{d)} +  \frac{1}{4}V_a^{(c} V_b^{d)}. \nn
\eea
Here also the auxiliary field was put to zero after calculation.

Identity \p{BIid5}, actually, should be brought to the canonical form $\partial_{ab}F^{ab} =0$. Unfortunately, this cannot be done unless one substitutes $X_{ab}$ as a function of $\partial_{ab}q$ and $V_{ab}$. In the third power in fields, it is just enough to take $X_{ab}$ in the lowest approximation, $X_{ab}\approx \partial_{ab}q$. Then after a few integrations \p{BIid5} can be shown to be equivalent to
\be\label{BIid6}
\partial_{ab}\Big[ V^{ab} + \frac{1}{8}V_{mn}V^{mn}\, V^{ab} + \frac{1}{4} \partial^{ac}q\, \partial^{bd}q\, V_{cd}  \Big]=0,
\ee
which assures that identity \p{BIid5} is benign and the right irreducibility conditions are chosen. However, if one needs to go further and find the complete field strength, it is needed to use the solution of equation \p{XBV4} with respect to $X_{ab}$. Note that solving \p{XBV4} for $\partial_{ab}q$ to find identity \p{BIid5} in terms of $V_{ab}$, $X_{ab}$ would not do the job: in the third power in fields, it is required to take into account that $\partial_{cd}X_{ab}\approx \partial_{ab}X_{cd}$ to bring the identity to the desired form.

Let us mention that in the systems with higher supersymmetries, such as the $N=4$, $d=4$ Born-Infeld theory, the difficulties of solving the relations between the components and calculation of the Bianchi identities, are even more severe \cite{N4d4BI}. Therefore, another approach to such systems should be devised.

\section{Alternative description}
To avoid the need to solve the quadratic equation and to find Bianchi identity in terms of two, not three ($V_{ab}$, $\partial_{ab}q$, $X_{ab}$) vectors, one may try to find the formulation of the multiplet in terms of the fermionic fields only. Indeed, in such formulation, one deals only with $X_{ab}$ and $V_{ab}$ as bosonic components, and any sort of Bianchi identity will be found without $\partial_{ab}q$. Also, there is intrinsically no quadratic equations on components, instead, in this formulation some identity should appear that implies that $X_{ab}$ is related to a derivative of a scalar. This identity should be some generalization of $\partial_{ab}X_{cd}- \partial_{cd}X_{ab}\approx 0$. Therefore, it may be used for simplification of Bianchi identity for the field strength, without need to solve this identity explicitly.

The irreducibility conditions \p{3dvecirr4} were obviously written in terms of fermionic superfields \p{3dvecirr4}, but it would be impossible to derive the Bianchi identity from them without taking into account that $\mpsi_a = \frac{\im}{2} \nabla_a \mq$. Another condition on $\bnabla_a \mpsi_b$ is, therefore, required. To find it, one may note that the relation in the linear limit \p{3dvecirr4} implies
\be\label{3dvecirr5}
D_a \mbpsi_b + \bD_b \mpsi_a  = \partial_{ab} \mq \; \Rightarrow  \; D_a \mbpsi^a + \bD^a \mpsi_a =0.
\ee
This relation works well in this limit: upon action of $D_a \bD_b$ with the use of only $D_a \mpsi_b=0$, $\bD_a \mbpsi_b=0$ it produces the condition
\be\label{newirr1}
\partial_{b}^{c} D_{(a} \mbpsi_{c)} + \partial_{a}^c \bD_{(b} \mpsi_{c)} =0
\ee
(antisymmetric parts of $\bD_a \mpsi_c$ are combined into a derivative of the original identity $\partial_{ab}\big( D_c \mbpsi^c + \bD^c \mpsi_c  \big)=0$). Substituting here the real and imaginary parts of $D_a \mbpsi_b$ just as in \p{XBV}, one may immediately find
\bea\label{XVeq1}
\partial_{bm}X_a^m + \partial_{am}X^m_b + \im \epsilon_{ab} \partial_{mn}V^{mn}=0,
\eea
Therefore, $V^{ab}$ satisfies the identity $\partial_{ab}V^{ab}=0$ and appears to be the electromagnetic field strength in $d=3$. For $X_{ab}$, the identity in vector notation is $\epsilon^{ABC}\partial_A X_B=0$, and $X_A$ is just a derivative of a scalar, as it should be \footnote{The transfer to vector notation can be performed with the help of three real symmetric matrices $\big(\sigma^A\big)_{ab}$, $\big(\sigma^A\big)_{ab}\big(\sigma^B\big)^{bc} =\delta_a^c \eta^{AB} + \epsilon^{ABC}\big(\sigma_C\big)_{a}^c$, $\eta^{AB} = \mbox{diag}(1,-1,-1)$.}.

Condition \p{3dvecirr5} looks very much like one of the $N=1$, $d=4$ vector supermultiplet irreducibility conditions \cite{BagGalvec} and can be obtained by dimensional reduction from it. Therefore, the full non-linear generalization of \p{3dvecirr5} should be sought with the use of the idea, which allowed us to derive the nonlinear irreducibility condition for the $N=2$, $d=4$ Born-Infeld theory \cite{N2d4BI}.

The required irreducibility condition can be constructed as follows. At first, one should introduce the scalar automorphism generator of algebra \p{N8d3Poincare1}, \p{N8d3Poincare2}, that mixes $Q_a$ and $\bS_a$, and put it in the coset with the auxiliary field as a parameter. Then the irreducibility conditions can be found as projections of the resulting extended $\omega_S$ forms, with respect to new $\omega_Q$ forms. Indeed, the superalgebra \p{N8d3Poincare1}, \p{N8d3Poincare2} can be extended with the automorphism generator with commutation relations
\be\label{Ucomms}
\big[ U, Q_a   \big] =\im \bS_a, \; \big[ U, S_a   \big] =-\im \bQ_a, \; \big[ U, \bQ_a   \big] =\im S_a, \; \big[ U, \bS_a   \big] =-\im Q_a.
\ee
Let us note that it satisfies the Jacobi identities involving $K_{ab}$, which had no analog in the $N=2$, $d=4$ Born-Infeld theory.

Introducing a new coset
\be\label{cosetU}
g_U =  e^{\im x^{ab}P_{ab}} e^{\theta^a Q_{a} + \bar\theta^a \bQ_a} e^{\mpsi^a S_{a} + \mbpsi^a \bS_a} \, e^{\im \varphi U },
\ee
one may calculate the Cartan forms and find that fermionic forms are modified to be
\bea\label{formsU}
\big(\omega_Q \big)^a = \cos \varphi \, d\theta^a + \sin \varphi\, d\mbpsi^a, \; \big( \bar\omega_Q  \big)^a = \cos\varphi \,d\bar\theta^a+\sin \varphi\, d\mpsi^a, \nn \\
\big(\omega_S \big)^a = \cos \varphi \, d\mpsi^a - \sin\varphi\, d\bar\theta^a, \; \big( \bar\omega_S \big)^a = \cos\varphi\, d\mbpsi^a-\sin \varphi\, d\theta^a.
\eea
This makes possible to introduce the covariant derivatives with respect to the $\omega_Q$-forms
\bea\label{newders1}
\big(\omega_Q \big)^a \mD_a + \big( \bar\omega_Q \big)^a \mbD_a + \triangle x^{ab} \mD_{ab} =  \triangle x^{ab} \nabla_{ab} + d\theta^a \nabla_a + d\bar\theta^a \bnabla_a, \nn \\
\mD_a = \frac{1}{\cos\varphi}\nabla_a - \tan\varphi \big( \mD_a \mbpsi^b\, \nabla_b + \mD_a \mpsi^b\, \bnabla_b   \big),\\
\mbD_a = \frac{1}{\cos\varphi}\bnabla_a - \tan\varphi \big( \mbD_a \mbpsi^b\, \nabla_b + \mbD_a \mpsi^b\, \bnabla_b   \big).\nn
\eea
The generalization of conditions  \p{3dvecirr4}, \p{3dvecirr5} can be found, if one puts to zero the $\omega_Q$ projection of $\omega_S$, as well as trace of its $\bar\omega_Q$ projection. This results in the conditions
\be\label{newirr}
\mD_a \mpsi_b =0, \; \mbD_a \mbpsi_b =0, \; \mbD_a \mpsi^a =- \mD^a \mbpsi_a = 2\sin\varphi.
\ee
To use these conditions in actual calculations, one should formulate them in terms of more convenient $\nabla_a$, $\bnabla_b$ derivatives and exclude the field $\varphi$. This calculation is analogous to the simplification of constraints of the $N=2$, $d=4$ Born-Infeld theory \cite{N2d4BI} and can be performed with the use of similar ideas. We, therefore, omit it and present only the final result:
\be\label{newirr2}
\nabla_a \mpsi_b =0, \; \bnabla_a \mbpsi_b =0, \; \mbox{and} \;
\big( 1 - 1/2\, \bnabla^{m}\mpsi^n \, \bnabla_m \mpsi_n   \big)\nabla_a \mbpsi^a = \big( 1 - 1/2\, \nabla^{m}\mbpsi{}^n \, \nabla_m \mbpsi_n   \big)\bnabla_a \mpsi^a.
\ee
To derive the full nonlinear Bianchi identities for $V_{ab}$, $X_{ab}$, one needs to apply $\nabla_a \, \bnabla_b$ to the second of equations \p{newirr2}. Putting to zero the auxiliary field after differentiation and taking the bosonic limit, one may find the identity
\be\label{newBI1}
\frac{\im}{2}\big( 1-1/2 \, \bnabla^{m}\mpsi^n \, \bnabla_m \mpsi_n   \big)\big\{ \nabla_c,\bnabla_b  \big\}\nabla_a \mbpsi^c + \frac{\im}{2}\big( 1 - 1/2\, \nabla^{m}\mbpsi{}^n \, \nabla_m \mbpsi_n   \big) \big\{ \nabla_a,\bnabla_c  \big\}\bnabla^c\mpsi_b=0.
\ee
Substituting here the on-shell expressions for the components of the fermionic superfields \p{XBV},
one may find that the imaginary part, which begins with the derivative of $V_{ab}$, is automatically antisymmetric in $(a,b)$, while the real part is automatically symmetric. It is more convenient to write them in the vector notation,
\bea\label{newBI2}
\big( \mbox{BI}_{(1)} \big) &=&\Big( 1 - \frac{1}{4}X^2 + \frac{1}{4}V^2   \Big)M_A{}^B \partial_B V^A + \frac{1}{2}\big(XV \big)M_{A}{}^B \partial_B X^A =0, \ \\
\label{newBI3}\big( \mbox{BI}_{(2)} \big)^A &=& \Big( 1 - \frac{1}{4}X^2 + \frac{1}{4}V^2   \Big) \Big( \epsilon^{ABC}M_B{}^D\partial_D X_C + \frac{1}{2}\,\epsilon^{MNP}\,\partial_M V^A\, X_N\, V_P  \Big) - \\&&- \frac{1}{2}\big( XV \big)\Big( \epsilon^{ABC}M_B{}^D\partial_D V_C - \frac{1}{2}\, \epsilon^{MNP}\,\partial_M X^A\, X_N\, V_P    \Big)=0,\nn \\
M_{A}{}^B &=& \big( 1 - \frac{1}{4}V^2 - \frac{1}{4}X^2 \big)\delta_A^B + \frac{1}{2}X_A X^B + \frac{1}{2}V_A V^B, \; X^2 = X_A X^A. \nn
\eea
As expected, in the lowest approximation they imply $\partial_{A}V^{A} \approx 0$ and $\epsilon^{ABC}\partial_B X_C \approx 0$, and $X_A$ can be reduced to a function of a derivative of some scalar $\partial_A q$ and $V_A$.

It is worth noting that identity \p{newBI3} is actually equivalent to quadratic equation \p{XBV4}. The simplest way to prove this is to solve equation \p{XBV4} for $\partial_A q$ in the bosonic limit
\be\label{dqXV}
\partial_A q = \frac{\Big( 1 - \frac{1}{4}X^2 +\frac{1}{4}V^2  \Big)X_A - \frac{1}{2}(XV)V_A}{ 1 - \frac{1}{16}\big( X^2 -V^2  \big)^2 - \frac{1}{4}(XV)^2}.
\ee
For the left-hand side it, obviously, should be $\epsilon^{ABC}\partial_B \partial_C q=0$. The same is true for the right-hand side if one takes into account both identities \p{newBI2} and \p{newBI3}.

Let us now find the physical field strength in the bosonic limit. As it is easy to find, $\big( \mbox{BI}_{(1)} \big)$ \p{newBI2} cannot be brought to the standard form $\partial_A F^A=0$ on its own, and  it is necessary to study a combination of \p{newBI2} and \p{newBI3}:
\be\label{newBI6}
\mu \cdot \big( \mbox{BI}_{(1)} \big) + \nu \cdot \epsilon^{ABC}X_B \, V_C \big( \mbox{BI}_{(2)} \big)_A = \partial_A \big(   f \, V^A + g \, X^A\big),
\ee
where $\mu$, $\nu$, $f$, $g$ are the functions of $X^2 = X_A \, X^A$, $V^2= V_A\, V^A$, $XV = X_A \, V^A$.
Substituting the matrix $M_A{}^B$ explicitly, one may extract ten equations, these functions should obey, as coefficients of linearly independent combinations $\partial_A V^A$, $V^A\, X^B \, \partial_A V_B$, $\ldots$. A careful analysis shows that three functions can be algebraically expressed in terms of one:
\bea\label{fgmunu}
\mu = 2\nu \Big( 1 + \frac{1}{4}X^2 +\frac{1}{4}V^2  \Big), \nn \\
f = 2\nu \Big( 1 - \frac{1}{4}X^2 +\frac{1}{4}V^2  \Big) \Big[ 1 - \frac{1}{16}\big( X^2 -V^2  \big)^2 - \frac{1}{4}(XV)^2 \Big],\\
g = \nu (XV) \Big[ 1 - \frac{1}{16}\big( X^2 -V^2  \big)^2 - \frac{1}{4}(XV)^2 \Big]. \nn
\eea
Then the remaining differential equations imply
\be\label{nu}
\nu = \frac{\mbox{const}}{\Big[ 1 - \frac{1}{16}\big( X^2 -V^2  \big)^2 - \frac{1}{4}(XV)^2 \Big]^2}.
\ee
The physical field strength in the bosonic limit is, therefore, given by the equation
\be\label{Ftrue}
F^A = \frac{\Big( 1 - \frac{1}{4}X^2 +\frac{1}{4}V^2  \Big)V^A + \frac{1}{2}(XV)X^A}{ 1 - \frac{1}{16}\big( X^2 -V^2  \big)^2 - \frac{1}{4}(XV)^2}.
\ee

The last step in deriving the Bianchi identities is the restoration of the fermionic terms. Indeed, the Bianchi identities can be calculated with fermions taken into account. In the first approximation in bosons, the identity for $X_A$ after a very long calculation can be found to be
\bea\label{BIferm1}
\cD_{ac}X^c_b + \cD_{bc}X^c_a -\im \big( \cD_{ak}\psi^m \, \cD_{b}^k \bpsi^n +\cD_{bk}\psi^m \, \cD_{a}^k \bpsi^n \big)X_{mn}=0 \Rightarrow \nn \\
\epsilon^{ABC}\cD_A X_B - 2\im \epsilon^{ABC}\cD_A \psi^m \, \cD_B \bpsi^n\, \big(  \sigma^M \big)_{mn}X_M \Rightarrow \epsilon^{MNP}\partial_M \big( \cE_N{}^K \,X_K  \big)=0.
\eea
Therefore, in this approximation $X_A \approx \cD_A q$, and no new terms proportional to $V_A$ and fermions arise. It can be generalized with the help of the bosonic limit \p{dqXV} as
\be\label{dqXV2}
\cD_A q = \frac{\Big( 1 - \frac{1}{4}X^2 +\frac{1}{4}V^2  \Big)X_A - \frac{1}{2}(XV)V_A}{ 1 - \frac{1}{16}\big( X^2 -V^2  \big)^2 - \frac{1}{4}(XV)^2}.
\ee
The same, obviously, follows from equation \p{XBV4}.

In the same approximation in bosons, the identity for the field strength is
\be\label{BIferm2}
\cD_{ab}V^{ab} +2\im V^{mn} \big(\cD_{am}\,\psi_n \, \cD^a_c \bpsi^c - \cD^a_c \psi^c \, \cD_{am}\bpsi_n \big) +2 X^{ab}\,\cD_a^k \psi_m\, \cD_{bk}\bpsi^m=0.
\ee
After some calculations it can be found equivalent to
\be\label{BIferm3}
\partial_A \big\{ \det\cE \big( \cE^{-1}  \big)_B{}^A \big[V^A + \epsilon^{BCD}\cD_C q\, \big( \psi_m \, \cD_D \bpsi^m + \bpsi^m \, \cD_D \psi_m  \big)  \big]    \big\} =0,
\ee
with the nontrivial term containing $\cD_A q$ and fermions (to prove this, one has to explicitly use $X_A \approx \cD_A q$). Its generalization with the help of the bosonic limit \p{Ftrue} is not trivial. It can be argued that it should be generalized to
\be\label{BIferm4}
\partial_A \big\{ \det\cE \big( \cE^{-1}  \big)_B{}^A \big[F^A + \epsilon^{BCD}\cD_C q\, \big( \psi_m \, \cD_D \bpsi^m + \bpsi^m \, \cD_D \psi_m  \big)  \big]    \big\},
\ee
by just replacing $V_A \rightarrow F_A$, with the use of the idea of invariance with respect to broken supersymmetry. Indeed, a variation of the first part of \p{BIferm4} under the broken supersymmetry transformations
\be\label{bsusytr}
\delta_S \psi^a = \varepsilon^a, \; \delta_S x^A = \im \big( \varepsilon^a \bpsi^b + \bar\varepsilon^a \psi^b  \big)\big( \sigma^A \big)_{ab}
\ee
is proportional to itself
\be\label{bsusydf}
\delta_S \partial_A \big\{ \det\cE \big( \cE^{-1}  \big)_B{}^A F^A   \big\} = - \im \big( \varepsilon^a \partial_M\bpsi^b + \bar\varepsilon^a \partial_M \psi^b  \big)\big( \sigma^M \big)_{ab}\,\partial_A \big\{ \det\cE \big( \cE^{-1}  \big)_B{}^A F^A   \big\},
\ee
For the same to happen with the second half of \p{BIferm4}, $\cD_A q$ should be kept during generalization, otherwise shifts of the explicitly contained fermions would ruin the symmetry:
\bea\label{bsusydf2}
&&\delta_S \partial_A\big\{ \det\cE \big( \cE^{-1}  \big)_B{}^A \epsilon^{BCD}\cD_C q\, \big( \psi_m \, \cD_D \bpsi^m + \bpsi^m \, \cD_D \psi_m  \big)     \big\} = \nn \\
&&= - \im \big( \varepsilon^a \partial_M\bpsi^b + \bar\varepsilon^a \partial_M \psi^b  \big)\big( \sigma^M \big)_{ab}\,\partial_A\big\{ \det\cE \big( \cE^{-1}  \big)_B{}^A \epsilon^{BCD}\cD_C q\, \big( \psi_m \, \cD_D \bpsi^m + \bpsi^m \, \cD_D \psi_m  \big)     \big\} +\nn \\
&&+\partial_A\big\{ \det\cE \big( \cE^{-1}  \big)_B{}^A \epsilon^{BCD}\cD_C q\, \big( \varepsilon_m \, \cD_D \bpsi^m + \bar\varepsilon^m \, \cD_D \psi_m  \big)     \big\}.
\eea
If and only if $\cD_A q$ is present, $\cE_A{}^B$ can be removed from the last term, and then it would vanish due to the antisymmetric nature of $\epsilon^{ABC}$.

\section{The action and its unbroken supersymmetry invariance}
The action of the $N=4$, $d=3$ Born-Infeld theory is dual to the action of supermembrane in $D=5$. The dualization of one scalar field was performed in \cite{D5membr} with the result
\bea\label{action2}
S &=& -\int d^3 x\det\cE \Big[ 1+ \sqrt{1 - \cD_A q \cD^A q + F_A F^A - (F^B \cD_B q)^2 }  \Big] = \nn \\&&= -\int d^3 x\det\cE \Big[1 + \sqrt{\det\big( \eta_{AB}-\cD_A q\, \cD_B q + \epsilon_{ABC}F^C  \big)}\Big].
\eea
(with some redefinition of the fields).

Let us prove that this action is actually invariant with respect to both supersymmetries. While invariance with respect to the broken supersymmetry in almost obvious, as the integration measure and the derivatives are properly covariantized, the unbroken supersymmetry deserves more detailed discussion.

With the use of expressions \p{Ftrue}, \p{dqXV}, the action \p{action2} can be written in terms of $X^A$, $V^A$ or, most conveniently, $Y_A = X_A + \im V_A$:
\bea\label{action3}
S &=&  -\int d^3 x \det\cE \frac{2+ \frac{1}{2}\big( V^2-X^2 \big)}{1 - \frac{1}{16}\big(V^2-X^2\big)^2 - \frac{1}{4}(XV)^2} = - \int d^3 x \,\det\cE \frac{2- \frac{1}{4}\big( Y^2+\bY{}^2 \big)}{1 - \frac{1}{16}Y^2\bY{}^2}.
\eea
Another advantage of using such variables is that they appear everywhere in the unbroken supersymmetry transformation laws. Indeed, these laws can be found using the formula
\bea\label{Qsusytr}
&&\delta^\star_Q f = - \big( \epsilon^a D_a + \bar\epsilon^a \bD_a )\mf|_{\theta\rightarrow 0} =  - \big( \epsilon^a \nabla_a + \bar\epsilon^a \bnabla_a )\mf|_{\theta\rightarrow 0}- H^A \partial_A f, \nn \\
&&H^A = -\frac{\im}{2}\big( \epsilon_k \, \psi^k\, Y^A + \bar\epsilon_k \, \bpsi^k\, \bY^A \big) - \frac{\im}{2}\epsilon^{ABC}\big( \epsilon_k \, \psi^m \, Y_B + \bar\epsilon_k \, \bpsi^m \, \bY_B   \big)\big( \sigma_C \big)^k_m.
\eea
Then for all essential combinations of components the $\epsilon$-part of these laws reads
\bea\label{deltaQpsi}
\delta^\star_Q \det\cE &=&- \partial_A \big( \det\cE \, H^A  \big) - \im \det\cE \epsilon_k \, \cD_A\psi^k\, Y^A - \im \det\cE \epsilon^{ABC}\epsilon_k \, \cD_A \psi^m \, Y_B \big( \sigma_C \big)^k_m, \nn \\
\delta^\star_Q Y^2 &=& -4\im \frac{1-1/4 \, Y^2}{1-1/4\, \bY{}^2} \big[ \epsilon_m\, \cD_A \psi^m\, Y^A + \epsilon_m\, \epsilon^{ABC} \cD_A \psi^n \, Y_B\, \big( \sigma_C \big)^m_n   \big] -  \\
&&- Y^2 \frac{1-1/4\, Y^2}{1-1/4\, \bY{}^2} \big[ \epsilon_m\, \cD_A \psi^m\, \bY^A - \epsilon_m\, \epsilon^{ABC} \cD_A \psi^n \, \bY_B\, \big( \sigma_C \big)^m_n   \big] - H^C \partial_C Y^2,\nn \\
\delta^\star_Q \bY^2 &=& - 4\im \epsilon_m\, \cD_A \psi^m\, \bY^A + 4\im \epsilon_m\,\epsilon^{ABC}\, \cD_A \psi^n \bY_B \, \big( \sigma_C \big)^m_n -\nn \\
&&- \im \bY^2 \cdot \epsilon_m\, \cD_A \psi^m\, Y^A - \im \bY^2 \cdot \epsilon_m\, \epsilon^{ABC} \cD_A \psi^n \, Y_B\,\big( \sigma_C \big)^m_n - H^C \partial_C \bY^2.\nn
\eea
Variation of Lagrangian in \p{action3} under these transformations can be reduced just to
\be\label{deltaQlagr1}
\delta^\star_Q {\cal L} = -2 \det\cE \, \epsilon_m \, \cD_A \psi^m F^A +2 \im \det\cE \, \epsilon_m \, \epsilon^{ABC} \cD_A \psi^n \cD_B q\big( \sigma_C  \big)^m_n,
\ee
as due to \p{dqXV} and \p{Ftrue}
\be\label{dqVY}
\cD_A q + \im F_A = Y_A \frac{1- \frac{1}{4}\bY{}^2}{1- \frac{1}{16}Y^2\bY{}^2}.
\ee
In the first approximation in fermions the second term in \p{deltaQlagr1} reduces to the full derivative, while the first one, after integration by parts, produces the bosonic Bianchi identity $\partial_A F^A\approx 0$. A careful treatment of \p{deltaQlagr1} with the help of \p{BIferm4} and explicit substitution of $\cE_A{}^B$ show that fermionic terms also reduce to the full derivatives. The action \p{action2} is, therefore, invariant.

\section{Conclusion}
In this paper, we constructed, from first principles, the component action of the $N=4$, $d=3$ Born-Infeld theory and found physical field strength of the corresponding vector supermultiplet. To perform this calculation, it was found to be necessary to carefully choose constraints, which define the multiplet, as most straightforward constraints result in unacceptably complicated relations between components of the multiplet, as well as the Bianchi identity for the field strength. To avoid these difficulties, the formulation of the multiplet, in which the fermionic superfield has fundamental role, was used. To ensure it has the proper component content, an additional irreducibility condition was provided. This condition, after applying covariant derivatives to it, produces the differential identities obeyed by the bosonic components of the multiplet. These identities finally allow identification of these components as the electromagnetic field strength and a derivative of a scalar.

Rather similar difficulties, solution of relation between the multiplet components and calculation of the Bianchi identity, can be found when considering the $N=4$, $d=4$ Born-Infeld theory. It is probable that they can be solved in a similar way by providing proper irreducibility conditions of the nonlinear $N=2$, $d=4$ vector multiplet.

\section*{Acknowledgments}
The work of N.K. and S.K. was partially supported by the RSCF, grant 14-11-00598 and partially by the RFBR, grant 15-52-05022 Arm-a.

\end{document}